\documentclass[aps, prx, superscriptaddress, showpacs, twocolumn, reprint, longbibliography]{revtex4-1}

\usepackage{graphicx}
\usepackage{amssymb}
\usepackage{amsmath}
\usepackage{comment}
\usepackage{natbib}
\usepackage{braket}
\usepackage{makecell}
\usepackage{hyperref}
\usepackage{url}
\usepackage{pifont} 

\usepackage{soul} 
\usepackage{color}

\newcommand{\fbar}{\bar{F}}

\newcommand{\cp}{$\mathcal{P}$}
\newcommand{\ct}{$\mathcal{T}$}
\newcommand{\Q}{$Q$}

\begin{document}

\sethlcolor{yellow}
\soulregister\ref{7}
\soulregister\eqref{7}
\soulregister\cite{7}
\soulregister\onlinecite{7}

\renewcommand{\hl}[1]{#1}
\renewcommand{\st}[1]{}

\title{Reflectionless excitation of arbitrary photonic structures: \\ A general theory}
\author{A. Douglas Stone}
\email{douglas.stone@yale.edu}

\affiliation{Department of Applied Physics, Yale University, New Haven, CT 06520, USA}
\affiliation{Yale Quantum Institute, Yale University, New Haven, CT 06520, USA}

\author{William R. Sweeney}
\affiliation{Department of Physics, Yale University, New Haven, CT 06520, USA}

\author{Chia Wei Hsu}
\affiliation{Department of Applied Physics, Yale University, New Haven, CT 06520, USA}
\affiliation{Ming Hsieh Department of Electrical and Computer Engineering, University of Southern California, Los Angeles, California 90089, USA}

\author{Kabish Wisal}
\affiliation{Department of Physics, Yale University, New Haven, CT 06520, USA}

\author{Zeyu Wang}
\affiliation{Ming Hsieh Department of Electrical and Computer Engineering, University of Southern California, Los Angeles, California 90089, USA}

\date{\today}

\begin{abstract}
We outline and interpret a recently developed theory of impedance-matching, or reflectionless excitation of arbitrary finite photonic structures in any dimension.
The theory includes both the case of guided wave and free-space excitation.
It describes the necessary and sufficient conditions for perfectly reflectionless excitation to be possible, and specifies how many physical parameters must be tuned to achieve this.
In the absence of geometric symmetries, such as parity and time-reversal, the product of parity and time-reversal, or rotational symmetry, the tuning of at least one structural parameter will be necessary to achieve reflectionless excitation.
The theory employs a recently identified set of complex-frequency solutions of the Maxwell equations as a starting point, which are defined by having zero reflection into a chosen set of input channels, and which are referred to as R-zeros.
Tuning is generically necessary in order  to move an R-zero to the real-frequency axis, where it becomes a physical steady-state impedance-matched solution, which we refer to as a Reflectionless Scattering  Mode (RSM).
In addition, except in single-channel systems, the RSM corresponds to a particular input wavefront, and any other wavefront will generally not 
be reflectionless.
It is useful to consider the theory as representing a generalization of the concept of critical coupling of a resonator, but it holds in arbitrary dimension, for arbitrary number of channels, and even when resonances are not spectrally isolated.
In a structure with parity and time-reversal symmmetry (a real dielectric function) or with parity-time symmetry, generically a subset of the R-zeros have real frequencies, and reflectionless states exist at discrete frequencies without tuning.
However they do not exist within certain spectral ranges, as they do in the special case of the Fabry-P{\'e}rot or two-mirror resonator, due to a spontaneous symmetry-breaking phenomena when two RSMs meet.
Such symmetry-breaking transitions correspond to a new kind of exceptional point, only recently identified, at which the shape of the reflection and transmission resonance lineshape is flattened.
Numerical examples of RSMs are given for one-dimensional multi-mirror cavities, a two-dimensional multiwaveguide junction, and a multimode waveguide functioning as a perfect mode converter.
Two solution methods to find R-zeros and RSMs are discussed. 
The first one is a straightforward generalization of the complex scaling or perfectly matched layer (PML) method, and is applicable in a number of important cases; the second one involves a mode-specific boundary matching method that has only recently been demonstrated, and can be applied to all geometries for which the theory is valid, including free space and multimode waveguide problems of the type solved here.

\end{abstract}


\maketitle
\section{Introduction}

    \subsection{Reflectionless excitation of resonant structures}

Reflectionless excitation or transmission of waves is a central aspect of harnessing waves for distribution or transduction of energy and information in many fields of applied science and engineering.
In the context of radio-frequency and microwave electronics and in acoustics, this is typically referred to as ``impedance-matching", whereas in optics and photonics the terms ``index-matching"  and ``critical coupling" are more frequently used, as well as ``perfect absorption" when the goal is energy transfer or transduction.
In the first fields listed it is typical to represent the response of the media or circuits, which are typically lossy, via a complex impedance, and the simple principle of matching the input impedance to the output impedance is often employed to achieve reflectionless excitation.
In optics and photonics it is more typical to represent the response of the medium by a complex dielectric function or susceptibility, and nearly lossless excitation of dielectric media, as well as free-space excitation, is quite common, so the term impedance-matching is less often used.
In this article we will focus on optical and photonic structures/devices and will use the term reflectionless excitation.
We will define below the concept of reflectionless scattering  modes (RSMs), referring to input wavefronts at specific, discrete {\it real} frequencies that can be shown to excite a given structure with zero reflection (in a sense to be clearly defined below).
Impedance-matching or index-matching across boundaries between effectively semi-infinite media is well-known from textbooks and is not the  topic of interest here.
Rather here we focus on {\it finite structures} in any dimension, which  are excited by a wave with wavelength smaller than the relevant dimensions of the structure.
In this case reflectionless excitation may be possible, but only at discrete frequencies, due to the necessity of taking into account multiple internal reflections within the structure.
Thus we  are  speaking of resonant reflectionless excitation of the structure.

The concept of critical coupling to a resonator, to be discussed in more detail below, is reasonably well known in optics and photonics: a high-\Q\ resonator, when excited by a single electromagnetic channel, either guided or radiative, will generate no reflected waves when it is excited at the resonance frequency and the input coupling rate to the resonator equals the sum of all {\it other} loss rates from or within the resonator.  
The total loss of the resonator is defined as the imaginary part of the complex frequency of the specific quasi-normal mode being excited, which includes the loss through the input channel.
The quasi-normal modes (or simply resonances) are rigorously defined as the purely outgoing solutions  of the relevant electromagnetic wave equation.
These resonances generically have frequencies in the lower half-plane,  $\omega = \omega_r -i\gamma$, where $\gamma = 1/2\tau > 0$, $\tau$ is the dwell time or intensity decay rate, and $Q=\omega_r \tau$ is the quality factor of the resonance~\cite{1928_Gamow_ZFP,1981_Bohm_JMP,1998_Ching_RMP,2018_Lalanne_LPR}.
In general, the resonances are not physically realizable steady-state solutions, due to their exponential growth at infinity, but they determine the scattering behavior under steady-state (real frequency) harmonic excitation.
However in electromagnetic scattering with gain, the resonances {\it can} be realized physically and correspond to the onset of laser emission~\cite{1973_Lang_PRA,2010_Ge_PRA,2014_Esterhazy_PRA}.
Thus, in the terminology we use in this work, having a resonance on the real axis does not correspond to the existence of a reflectionless state  (in some other contexts the term ``resonance" is used to refer to a reflectionless state).
In the current work we will define a different set of complex-frequency solutions which  {\it do} correspond to the existence of a reflectionless state, and which do not in general require the addition of gain or loss to the system to make them accessible via steady-state excitation.

The current theoretical/computational tools available in optics and photonics for determining when and if reflectionless excitation of a structure/resonator is possible consists of analytic calculations in certain one-dimensional structures, and transfer matrix computations for more complicated one-dimensional or quasi-one-dimensional structures, along with the principle of critical coupling, which rarely is applied in higher dimensions.

    \subsection{Limitations of critical coupling concept}

The terminology ``critical coupling" (CC) appears to have  been used in microwave/radio-frequency electronics at least sixty years ago but does not appear to have been used extensively in optics until the nineties~\cite{adler_1960_book,yan_1989_ieee,law_1990_2_apl,law_1990_apl}.
It always is applied to a structure in which a relatively high-\Q\ resonator with well separated resonances is effectively excited by a single spatial input channel.
The resonator will have some effective coupling-in rate at the surface where the input channel comes in, determined, e.g., by a mirror or facet reflectivity, and it will have some coupling-out/absorption rate within the resonator, due either to other radiative channels or to internal absorption loss or both~\cite{2000_Cai,yariv_2002_ieee}.  
Examples in photonics include the asymmetric Fabry-P{\'e}rot semiconductor devices developed in the eighties and nineties, which used the electro-optic effect to switch on and off absorption in the cavity, so as to toggle between a critically coupled condition and a weakly coupled condition \cite{yan_1989_ieee,law_1990_2_apl,law_1990_apl}.  
In this case the loss is primarily absorption and represents an irreversible transduction of the energy.
Another common, more recent set of examples are the ring resonators side-coupled to silicon waveguides which can be toggled by free-carrier injection between a critically-coupled and a weakly coupled state, which turns off and on the transmission through the waveguide~\cite{lipson_2005_JLWT}.
In this case the loss in the critically coupled state is primarily radiative, and the reflectionless ``on'' state can be thought of as perfect transmission into the radiative channels, whereas the ``off'' state corresponds to zero reflection from the ring (only) and hence continued propagation/transmission along the guided waveguide channel.  
These examples indicate that the source of the loss in the resonator is not important to be able to achieve critical coupling, although it does determine the effect of critical coupling on the exciting wave, i.e. either irreversible transduction or radiative transmission (to a receiver or just to an effective beam sink).

The power of the critical coupling (CC) concept is that, if rigorously correct, it implies that it is possible to excite a resonator of arbitrary complexity through a single input channel and have zero reflection, if the total loss from either absorption or radiation into other channels equals the input coupling.
To our knowledge, the CC concept is the only general principle relating to reflectionless excitation of a resonator which applies beyond parity-symmetric one-dimensional examples, where it is possible to calculate analytically the reflectionless input frequencies.

However there are obvious questions raised about the meaning and generality of the concept. 
\begin{itemize}

    \item In all cases of which we are aware, the CC condition is derived within a simplified coupled-mode theory, and not from a first-principles analysis, which would be exact within Maxwell electrodynamics.
    What principle, if any, underlies its validity in the case of a complex resonator with no symmetries?
    
    \item Is there a generalization of the CC concept to the situation in which one is exciting the resonator with more than one input channel?
    A simple example of this would be a multimode waveguide exciting a resonator or multiple-waveguide junction.
    Similarly, when one is exciting a structure larger than the excitation wavelength in free space, the typical radiation will involve higher multipoles and hence multiple input/output channels.
    Are there reflectionless solutions in either of these cases?
    
    \item Critical coupling assumes that the excitation is only of a single resonance, but in any relatively open structure, e.g., a waveguide junction, multiple resonances will often be overlapping and relevant to the scattering process.
    In this case the CC concept becomes ill-defined.
    There  is no obvious scalar meaning to coupling in and coupling out, so there is no obvious CC condition to apply.
    In fact the CC condition has never, to our knowledge, been applied to such situations and would generally be  considered  irrelevant because one doesn't have isolated resonances.

\end{itemize}

This paper outlines and interprets recent results from our group which answers all of these questions and proves that reflectionless states are an exact property of Maxwell electrodynamics in any dimension and for arbitrarily complex structures (larger than the exciting wavelength).
Moreover these states exist even when multiple resonances overlap and there are no isolated resonances.
In general a single continuous parameter of the resonator/structure needs to be tuned appropriately, and then the reflectionless state exists only at a single frequency.
The reflectionless states can be computed by numerical methods which are closely related to standard techniques in photonics and are tractable for realistic structures.
Hence we believe that the theory presented here provides a powerful tool for the design of photonic structures with controlled excitation which implement perfect coupling, as well as clarifying what sorts of solutions exist generically and what sorts do not.
We will present here only the main results of our analysis with illustrative examples; the detailed derivations are given in Ref.~\cite{rsm_arxiv}.

\section{The Generalized Reflection Matrix, R-zeros and RSMs\label{sec:intro to S-matrix}}

    \subsection{The scattering matrix}

To define reflectionless states in electromagnetic scattering we must first define the scattering matrix (S-matrix) of a finite photonic structure. 
We consider here the most general system
of interest, which consists of an inhomogeneous scattering region or structure, outside of which are asymptotic regions that extend to infinity.
\hl{To support resonance effects the scattering region needs to be larger than the wavelength of the excitations created by the input waves within the scatterer.
For dielectric systems this is typically of order, but smaller than, the input wavelength.
However the theory will also apply to metallic/plasmonic systems (within the Maxwellian framework), where the plasmonic excitations can have orders of magnitude smaller wavelengths.
An example of an application of Coherent Perfect Absorption, a special case of the theory, to nanoparticles is cited in the next section.}
The asymptotic regions are assumed to be time-reversal invariant (so that they support incoming and outgoing asymptotic channels that are related by complex conjugation) and to have some form of translational invariance, e.g., vacuum or uniform dielectric, or a finite set of waveguides, or an infinite periodic photonic crystal. The theory will apply to both free and guided waves. 
We also focus on media in which the scattering forces are short-range, i.e. net neutral media, which is typical for most photonic structures.

A linear and static photonic structure is described by its dielectric function $\varepsilon({\bf r,\omega})$, which is generically complex-valued, with its imaginary part describing absorption and/or gain. 
The assumed linearity allows the theory to concentrate on scattering at a single real frequency, $\omega$; time-dependent scattering can be studied by superposing solutions.
The translational symmetry of the asymptotic regions allows one to define $2N$ power-orthogonal propagating ``channel states'' at each $\omega$.
Based on the direction of their fluxes, the $2N$ channels can be unambiguously grouped into $N$ incoming and $N$ outgoing channels, which, as noted, are related by time-reversal.
Familiar examples of channels include the guided transverse modes of a waveguide and orbital angular-momentum waves in free space, with one channel per polarization.
In the waveguides, the finite number and width of the waveguides lead to a finite $N$ for a given $\omega$
, whereas for the case of a finite scatterer/cavity in free space the number of propagating angular-momentum channels is countably infinite.
However, a finite scatterer of linear scale $R$, with no long-range potential outside, will interact with only a finite number of angular momentum states, such that $l_{\rm max} \sim \sqrt{ \bar{\epsilon} }R \omega /c$, where $\bar{\epsilon}$ is the spatially-averaged dielectric function in the scattering region, and $c$ is the speed of light.
Hence for each $\omega$ we can reasonably truncate the infinite dimensional channel-space to a finite, $N$-dimensional subspace of relevant channels.

\begin{figure}[tb]
    \centering
    \includegraphics[width=0.85\columnwidth]{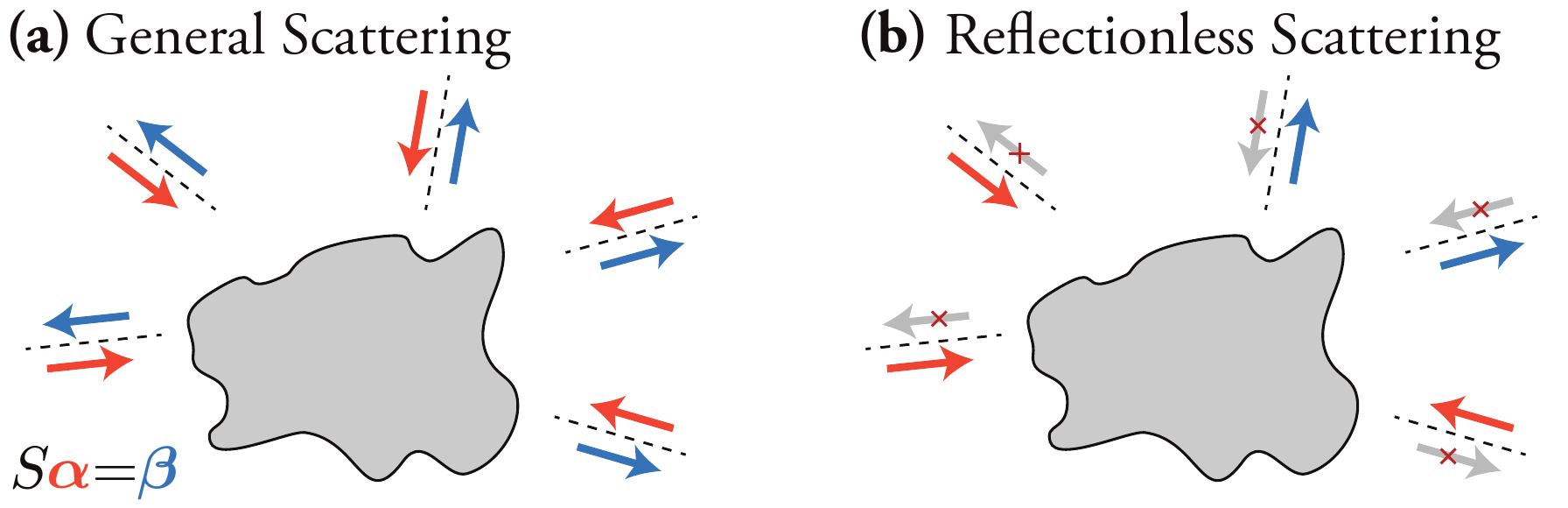}
    \caption{(Color) Schematic depicting a general scattering process (a) and reflectionless process (b).
    A finite scatterer/cavity interacts with a finite set of asymptotic incoming and  outgoing channels, indicated by the red and blue arrows, respectively, related by time-reversal.
    These channels may be localized in space (e.g., waveguide channels) or in momentum space (e.g., angular-momentum channels).
    {\bf (a)} In the general case without symmetry, all incoming channels will scatter into all outgoing channels. 
    {\bf (b)} There exist reflectionless states, for which there is no reflection back into a chosen set of incoming channels (the inputs), which in general occur at discrete complex frequencies and do not correspond to a steady-state harmonic solution of the wave equation.
    However, with variation of the cavity parameters, a solution can be tuned to have a real frequency, giving rise to a steady-state reflectionless scattering process for a specific coherent input state, referred to as a Reflectionless Scattering Mode (RSM).
    }
    \label{fig:schematic}
\end{figure}

A general scattering process then consists of incident radiation, propagating along the $N$ incoming channels, interacting with the scatterer and then propagating out to infinity along the $N$ outgoing channels, as illustrated in Fig.~\ref{fig:schematic}(a).
In a general geometry, which is not partitioned into spatially distinct asymptotic channels, there is no natural definition of reflection and transmission coefficients between different channels, only interchannel scattering vs. backscattering into the same channel; for certain geometries, such as a scattering region within a single or multi-mode fiber, it is natural to segment the S-matrix into reflection and transmission matrices depending on whether the scattering maintains the sign of propagation (i.e. transmits flux) or reverses it (reflects flux).
However we will define the reflection matrices for a general geometry in a more general way, which need not reduce to this standard definition even in a waveguide geometry.
In the channel basis, the wavefronts of the incoming and outgoing fields are given by length-$N$ column vectors ${\boldsymbol \alpha}$ and ${\boldsymbol \beta}$, normalized such that ${\boldsymbol \alpha}^\dagger {\boldsymbol \alpha}$ and ${\boldsymbol \beta}^\dagger {\boldsymbol \beta}$ are proportional to the total incoming and total outgoing energy flux, respectively.
The $N$-by-$N$ scattering matrix ${\bf S}(\omega)$, which relates ${\boldsymbol \alpha}$ and ${\boldsymbol \beta}$ at frequency $\omega$ is defined by
\begin{equation}
    \label{eq:S}
    {\boldsymbol \beta} = {\bf S}(\omega) {\boldsymbol \alpha}.
\end{equation}
In reciprocal systems, the S-matrix is symmetric, ${\bf S} = {\bf S}^T$~\cite{2013_Jalas_nphoton}.
If the scatterer is lossless (i.e., $\varepsilon$ is real everywhere), then any incoming state leads to a non-zero flux-conserving output, and the S-matrix is unitary.
However the theory outlined below is developed for arbitrary linear S-matrices and complex $\varepsilon$, which then includes the effects of linear absorption or amplification inside the scattering region. 
Engineering reflectionless states will generally be possible for the case of both unitary and non-unitary S-matrices.

    \subsection{Coherent Perfect Absorption}

The S-matrix, being well-defined at all real frequencies, can be extended to complex frequencies via analytic continuation.
As noted in the introduction, the resonances or quasi-normal modes of the system are solutions of the wave equation for the structure which are purely outgoing in the asymptotic regions and have discrete complex frequencies in the lower half-plane; those frequencies correspond to poles of the S-matrix.  
Since purely incoming solutions can be obtained by complex conjugation, this implies that the zeros (frequencies of solutions at which zero flux is outgoing) are simply the complex conjugate of the pole frequencies when the structure is lossless, and these  zeros occur in the upper half-plane.
The zeros then correspond to a certain type of reflectionless solution to the wave equation but at an unphysical complex frequency.
These states can be tuned to a real frequency by adding loss, in just the same manner as poles can be tuned to the real axis by adding gain to initiate lasing \cite{2010_Chong_PRL,2011_Wan_Science,2012_noh_prl,2017_Baranov_NRM}.
A system so tuned is known as a Coherent Perfect Absorber (CPA), and functions as the time-reverse of a laser at threshold.

\begin{figure}[tb!]
    \centering
    \includegraphics[width=0.85\columnwidth]{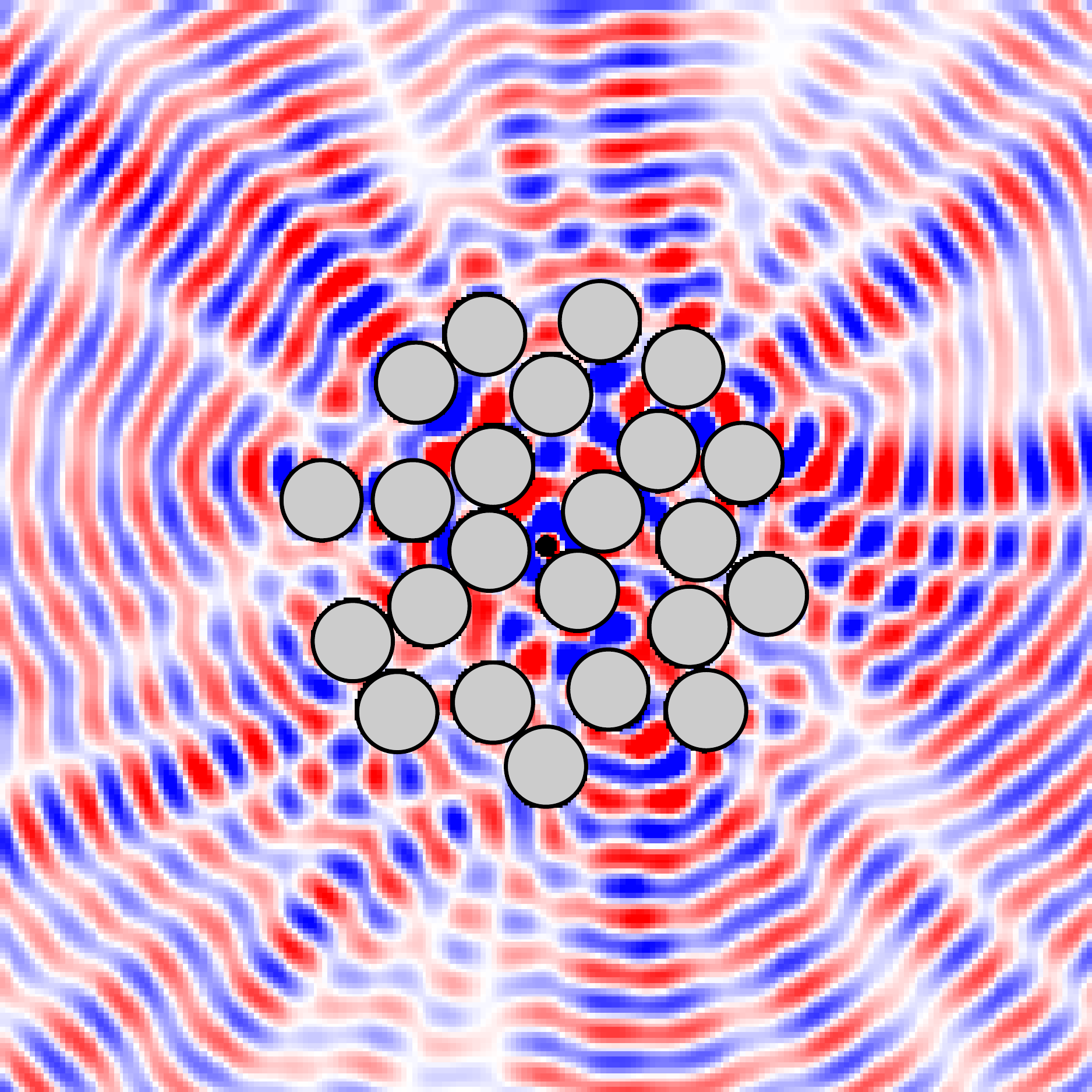}
    \caption{(Color) Reflectionless state of a coherent perfect absorber (CPA), consisting of a random aggregate of lossless glass scattering rods (gray) of index $n=1.5$ and radius equal to the incident wavelength ($r=\lambda$), surrounding a highly lossy subwavelength central rod (black) of radius $r=0.15\lambda$ and dielectric constant $\varepsilon= 1.28 +1.75i$.
    The color scale indicates the field amplitude for the specific input mode which is perfectly absorbed.
    This is a steady-state solution in which all of the incident power is dissipated in the central rod, which acts as a perfect sink and is assumed to be a linear absorber.
    The incident field pattern is found by calculating the complex conjugate of the threshold lasing mode of the analogous random laser, and consists of the appropriate coherent superposition of converging cylindrical waves (Hankel functions).
    The field penetrating into the glass rods is not shown, for clarity.
    In this very open structure the resonances strongly overlap and the concept of critical coupling to a single resonance doesn't apply, nonetheless reflectionless states exist.
    Figure adapted from animation at \href{http://www1.spms.ntu.edu.sg/~ydchong/research.html}{\url{http://www1.spms.ntu.edu.sg/~ydchong/research.html}}, courtesy of Y-D. Chong.
    \label{fig: randomCPA}
    }
\end{figure}

This type of of reflectionless state was first pointed out by one of the authors (and co-workers) a decade ago.
It is a special case of the reflectionless states we define below,  with the additional property that impedance-matching is achieved by perfect absorption within the scattering region, and hence irreversible transduction of the incident energy into degrees of freedom within the resonator/scatterer.
With the introduction of the CPA concept it was appreciated that this kind of reflectionless state always exists within a family of arbitrarily complex resonators with tunable loss, for just the same reason that any complex resonator can be made to lase with sufficient gain added.
And, just as the laser is ``perfectly emitting"  only for a specific spatial mode of the electromagnetic field, the CPA is perfectly absorbing only for a specific mode, which is the time-reverse of the lasing mode, and may have a very complicated spatial structure that is challenging to synthesize.
For example, just as there exist random lasers that emit a pseudo-random  lasing field when sufficient gain is added, a geometrically similar random structure with an absorbing medium added of similarly strong loss, can, in steady-state, perfectly absorb the complex conjugate of this lasing field (an example is shown in Fig.~\ref{fig: randomCPA}). 
\hl{As this example shows, in order to be perfectly absorbed the input ``beam'' must be focused to roughly the size of the scattering structure, so that CPA cannot be achieved in free space for an input beam (plane wave or structured) that is spread out over an area much larger than the transverse size of the structure.
Similar constraints apply to the more general reflectionless modes defined below, which will also need to be focused so as to strongly interact with the structure in free space (this constraint is typically automatically imposed by the geometry in the case of guided wave systems).}

However, as already noted, it can be quite difficult to synthesize the wavefront needed to achieve CPA, and this limits the application of the CPA concept to complex structures in experiments or devices.
In the more common situation in which one seeks to excite a structure in a reflectionless manner, one is not aiming for perfect absorption, but is simply seeking to avoid energy flow back into the chosen input channels; and in many cases one  doesn't wish to have any absorption at all.
Our theory below includes CPA as a limiting case but is focused instead on this more common situation, of prime importance for the design of photonic devices.


    \subsection{Generalized Reflection Matrix and R-zeros \label{sec:R-zeros}}

Returning to the S-matrix for an arbitrary finite scattering structure/resonator, we now define reflectionless states in the most general manner possible.
The full S-matrix encodes the information about all possible linear excitations of the resonator.  
Assuming one has access to all $N$ of the possible input channels in the asymptotic region, one can define a particular impedance-matching problem as shown in Fig.~\ref{fig:schematic}(b), by specifying $N_{\rm in}$ (with $0<N_{\rm in} \leq N$) of the incoming channels as the controlled input channels, which, for the appropriate input state, will carry incident flux but no outgoing flux.
Conversely the complementary set of $N_{\rm out}=N-N_{\rm in}$ outgoing channels will carry any outgoing flux. 
This flux can be less than, equal to, or greater than the incident flux, depending on whether the resonator is attenuating, lossless or amplifying.

Let us for convenience redefine the S-matrix so that for each choice of input channels the first $N_{\rm in}$ columns of the S-matrix represent the scattering of the chosen $N_{\rm in}$ input channels.
This implies that we will only consider scattering input vectors, ${\boldsymbol \alpha}$, which are non-zero for their first $N_{\rm in}$ components (henceforth we will refer to this as the input wavefront).
Conversely, for the input wavefront to be reflectionless, the output vector, ${\boldsymbol \beta}$, must have {\it its} first $N_{\rm in}$ components be equal to zero.
Thus we can define the upper left $N_{\rm in} \times N_{\rm in}$ block of this S-matrix to be a generalized reflection matrix, ${\bf R}_{\rm in}(\omega)$.
The condition then for the existence of a reflectionless input state at some frequency $\omega = \omega_{\rm RZ}$ is that this matrix have an eigenvector with eigenvalue zero; this eigenvector is the $N_{\rm in}$-component vector consisting of the non-zero components of ${\boldsymbol \alpha}$.
The frequencies at which a reflectionless state exists are thus determined by the complex scalar equation.
\begin{equation}
    \label{eq:det_Rin_zero}
    \det {\bf R}_{\rm in}(\omega_{\rm RZ}) = 0,
\end{equation}
and in principle the frequencies and input wavefronts for reflectionless states could be found by searching for the zeros of this determinant in the complex frequency plane.

However another approach is more fruitful.
That is to regard the reflectionless boundary conditions as defining a non-linear eigenvalue problem on the Maxwell wave operator for which $\{\omega_{\rm RZ}\}$ are the eigenvalues, and use standard methods for solving general nonlinear eigenproblems~\cite{1968_avner_acta_math,2000_Golub_JCAM,2010_Asakura_JIAM,2011_Su_SIAM,2012_Beyn_LA}.
To impose the appropriate boundary conditions a familiar method in photonics is the use of perfectly-match layers (PMLs), normally used for finding purely outgoing solutions (resonances), but also applicable here in some cases.
For the most general cases the PML approach is not applicable, but a modification of previous boundary matching methods can be used, as will be discussed briefly below.

Anticipating results which will be demonstrated below, similar to the resonances, we find that the reflectionless input wavefronts will only exist at discrete {\it complex} values of the frequency, $\omega_{\rm RZ}$; however these frequencies will differ from the resonance frequencies, and they represent the complex-valued spectrum of a wave operator with a different physical meaning.
We will refer to these frequencies and the associated wave solutions as R-zeros (reflection zeros).
Since complex frequency solutions do not represent physically realizable steady-state solutions, a critical step in solving the impedance-matching problem in generic cases will be to tune parameters of the dielectric function in the wave operator in order to move an R-zero to the real-frequency axis.
When an R-zero is tuned to the real axis, we will refer to the reflectionless physical state which results as a Reflectionless Scattering Mode (RSM), in analogy to the term ``lasing modes" or ``CPA mode"  which is used for these related but distinct electromagnetic eigenvalue problems.
We note that, like a resonance, an R-zero can be transiently realized with a time-dependent input, even when it occurs at a complex frequency \cite{2017_Baranov_Optica}.

A previous work, \cite{2018_Bonnet-BenDhia_PRSA}, has introduced the notion of the R-zero spectrum in the more limited context of waveguides, focusing on single-mode cases, and pointed  out its relevance to impedance-matching.
These authors did not discuss the possibility of tuning the wave operator to create a physical steady-state, but studied a  one-dimensional parity symmetric case for which the R-zeros can have real frequency due to symmetry.

    \subsection{The RSM concept}

As noted, an RSM is a steady-state wave solution at a real frequency that excites a structure/resonator through specific input channels such that there is zero reflection back into the chosen channels.
It can be specified asymptotically by its frequency and wavefront ${\boldsymbol \alpha}$ (where here we refer to the $N_{\rm in}$ non-zero components of ${\boldsymbol \alpha}$).
The absence of reflection for the RSM incident wavefront is due to interference: the reflection amplitude of each input channel $i$ destructively interferes with the interchannel scattering from all the other input channels, $(R_{\rm in})_{ii}{\alpha}_i+\sum_{j\neq i} (R_{\rm in})_{ij}{\alpha}_j=0$ (which is just a restatement of the fact that ${\boldsymbol \alpha}$ is an eigenvector of $\bf R_{\rm in}$ with eigenvalue zero).
However, the scattering (``transmission'') into the chosen  output channels is not obtained from solving this equation alone, and must be determined by solving the full scattering problem at $\omega_{\rm RZ}$, i.e. by calculating all the rows of the S-matrix at $\omega_{\rm RZ}$.

The case of $N_{\rm in}=N$ corresponds to CPA \cite{2010_Chong_PRL}, (${\bf R}_{\rm in}={\bf S})$, which, as noted, has been identified and studied for some time.
In this case the R-zeros are the zeros of the full S-matrix, which are now relatively familiar objects. CPA, as an example of an RSM, is rather special because it requires that all the asymptotic channels be controlled in order to have the possibility of achieving CPA, which is often not practical in complex geometries.
Also, only in the case of CPA is it necessary to violate flux conservation to create an RSM (i.e. by adding an absorbing term to the dielectric function).
Thus CPA is mainly of interest in cases where the goal is not impedance-matching, but rather transduction or sinking of energy.

    \subsection{Wave-operator Theory of Zeros of the S-matrix and Generalized Reflection Matrix \label{sec:rigorous S}}

It is both mathematically convenient and helpful for physical insight to consider the R-zero/RSM problem from the point of view of the underlying wave-operator with boundary conditions.
While Eq.~(\ref{eq:det_Rin_zero}) defines the frequencies and wavefronts for which reflectionless states exist for a fixed structure/resonator, solving it will not simultaneously yield the field everywhere within the structure. 
For the latter, one will need to solve the full Maxwell wave equations subject to the boundary conditions at infinity which follow from the frequency and incident wavefront.

The R-zero/RSM boundary conditions lead to a more constrained scattering problem than the standard scattering boundary conditions in which only the input wavefront is specified, but no constraint is placed on the output wavefront. 
For the standard problem, a solution exists at every real $\omega$, but it generically involves all of the outgoing channels in the asymptotic region.
In the RSM/R-zero problem only, $N-N_{in}$ of the asymptotic outgoing channels are allowed to appear, and $N_{in}$ of the input channels generically appear.  
Thus solutions are constructed from only $N$ of the $2N$ possible asymptotic free solutions, and are not guaranteed to exist at all frequencies.
As the the previous analysis leading to Eq.~(\ref{eq:det_Rin_zero}) has shown, having an R-zero requires a certain scattering operator to be non-invertible, which we don't expect to happen generically.

A more familiar situation, in which we impose similar boundary conditions, is in calculating the resonances, using only the $N$ outgoing channel functions at infinity.
As already noted, this calculation can be posed as an electromagnetic eigenvalue problem for which the eigenvalues are the frequencies at which solutions exist.
For the resonance problem it is well-known that for finite structures of the type considered here, there are an infinite number of solutions at discrete frequencies and that these frequencies correspond to the poles of various response functions including the S-matrix that we have introduced above.
In the absence of gain, by causality these poles are restricted to the lower half-plane.
We will now discuss a wave operator representation of the S-matrix and the implications for its zeros, in preparation for the general theory of R-zeros in the next section.
The results we present in the next two sections are based on derivations from Ref.~\cite{rsm_arxiv} and here we simply present the main results without proof, introducing the minimum number of mathematical details necessary to make these results comprehensible.
Readers interested only in the basic results and examples may skip to the summary in Section \ref{sec:general properties}.

To introduce our notation, consider a wave operator $\hat A(\omega)$ acting on state $\ket{\omega}$, which satisfies $\hat A(\omega)\ket{\omega}=0$.
For electromagnetic scattering, we may choose the quantity $\braket{{\bf r}|\omega}$ as the magnetic field, ${\bf H}({\bf r})$, under harmonic excitation at $\omega$.
The Maxwell operator at frequency $\omega$ is given by
\begin{equation}
\label{eq:Maxwell Operator}
    \braket{{\bf r^\prime}|\hat A(\omega)|{\bf r}}=\delta({\bf r-r^\prime})\left\{\left(\frac{\omega}{c}\right)^2-\nabla\times\left(\frac{1}{\varepsilon({\bf r},\omega)}
    \nabla\times\right)\right\}.
\end{equation}
Here we have given the full vector Maxwell operator, for which all of our results are valid, but in the more detailed analysis below we only apply the theory to effectively one- and two-dimensional cases in which a scalar Helmholtz-type equation describes the solutions for the appropriate polarization.

In order to express the scattering matrix in terms of its resonances, we divide the system into two regions: the finite, inhomogeneous scattering region $\Omega$ in the interior, and the exterior asymptotic region $\bar\Omega$ that extends to infinity, which possesses a translational invariance broken only by the boundary, $\partial\Omega$, between $\Omega$ and $\bar\Omega$.
We separate operator $\hat A(\omega)$ into three pieces,
\begin{equation}
    \hat A(\omega) = [ \hat A_0(\omega) \oplus \hat A_c(\omega)] + \hat V(\omega),
\end{equation}
with $\hat A_0(\omega)$ identical to $\hat A(\omega)$ on its domain $\Omega$ and $\hat A_c(\omega)$ is identical to $\hat A(\omega)$ on $\bar{\Omega}$.
$\hat V(\omega)$ represents the residual coupling between the two regions.

The closed-cavity wave operator $\hat A_0(\omega)$ on $\Omega$, which we do not assume  to be hermitian, has a discrete spectrum of the form $\hat A_0(\omega_\mu)\ket{\mu}=0$ with eigenvalues $\{\omega_\mu\}$.
The boundary conditions on $\hat A_0(\omega)$ can be chosen to be Neumann, but the effect of coupling terms will introduce a self-energy which will account for the actual continuity conditions at the boundary of $\Omega$.  
The matrix ${\bf A}_0(\omega)$ is naturally defined by its matrix elements
\begin{equation}
    \label{eq:A0}
    A_0(\omega)_{\mu \nu} = \braket{\mu|\hat A_0(\omega)|\nu}.
\end{equation}
The asymptotic wave operator $\hat A_c(\omega)$ on $\bar\Omega$ has a countable number of eigenfunctions at every real value of $\omega$; these are the propagating channel functions which satisfy $\hat A_c(\omega)\ket{\omega,n}=0$.
The operator $\hat V$ connects these closed and continuum states, and its off-diagonal block is represented by the matrix ${\bf W}(\omega)$,
\begin{equation}
    W(\omega)_{\mu n} = \braket{\mu|\hat V(\omega)|n,\omega}.
\end{equation}
While ${\bf W}(\omega)$ in general also has a contribution from evanescent channels in $\bar{\Omega}$, we will neglect the effect of such channels henceforth in the current discussion, as they don't change the central results qualitatively \cite{rsm_arxiv}.

With these definitions one can derive a general relation~\cite{1969_Mahaux_book,1997_Beenakker_RMP,2003_Viviescas_PRA,2017_Rotter_RMP} between the matrices ${\bf S}$, ${\bf A}_0$, and ${\bf W}$, originally developed in nuclear physics, which allows us to find the poles and zeros of ${\bf S}$ through its determinant:
\begin{equation}
\label{eq:det(S)}
    \det {\bf S}(\omega) = \frac{\det{\boldsymbol (}{\bf A}_0 (\omega)-{\boldsymbol \Delta}(\omega) - i{\boldsymbol \Gamma(\omega)}{\boldsymbol )}}{\det{\boldsymbol (}{\bf A}_0 (\omega)-{\boldsymbol \Delta}(\omega) + i {\boldsymbol \Gamma(\omega)} 
    {\boldsymbol )}}.
\end{equation}
The two positive-definite operators which appear here are ${\boldsymbol \Delta}$  and
$ {\boldsymbol \Gamma} \equiv \pi {\bf W} {\bf W}^\dagger$, which arise from the coupling operator and, roughly speaking, induce a real and imaginary shift of the eigenvalues of the ``closed" system to account for the openness of the system.
The operator ${\boldsymbol \Delta}$ can be expressed in terms of an integral over the ${\bf W} {\bf W}^\dagger$ matrix \cite{rsm_arxiv} and is of less interest in the current context; both operators are infinite dimensional matrices in the space of resonances of the system.
The S-matrix however is a finite dimensional matrix  in the truncated channel space (as noted above) and hence Eq.~\eqref{eq:det(S)} is not a simple identity of linear algebra: the left-hand side is the standard determinant of an $N$-by-$N$ square matrix, while the right-hand side is a ratio of functional determinants of differential operators on an infinite-dimensional Hilbert space (see Ref.~\cite{rsm_arxiv} for more details and relevant references).
Since the right-hand side is a ratio of determinants of infinite dimensional differential operators upon a finite domain (which will have a countably infinite set of complex eigenvalues which depend on $\omega$), this implies that $\det {\bf S}$ indeed has a countably infinite set of zeros and poles (corresponding to the vanishing of the numerator and denominator).  
When ${\bf A}_0$ is hermitian (the scatterer is lossless), then the operators in the numerator and denominator are hermitian conjugates and the poles and zeros come in complex conjugate pairs.

Since the operators in each determinant of the ratio do not commute, one cannot simply say that the eigenvalues of the full operators are the sum of the eigenvalues of each individual operator.
However, the more isolated the resonances of the systems are, the more useful is this heuristic interpretation of Eq.~\eqref{eq:det(S)}.
Thus, crudely speaking, the scattering resonances will occur at complex frequencies where the real part is given by the real part of the eigenvalue of the closed system containing the scatterer, shifted by the contribution from ${\boldsymbol \Delta}$, while its imaginary part must be negative, with a value $i (-\gamma_{\mu,\rm rad} - \gamma_{\mu,\rm int})$, where $\gamma_{\mu,\rm rad}>0$ is an eigenvalue of ${\bf \Gamma}$ and represents radiative loss, and $\gamma_{\mu,\rm int}$ comes from gain or loss in the resonator and can have either sign.
With the standard convention we have chosen here $\gamma_{\mu,\rm int}>0$ corresponds to absorption loss, so that adding absorption pushes the resonance away from the real axis and hence broadens it, as is familiar.  
Conversely the zeros of ${\bf S}$ have imaginary part $i (\gamma_{\mu,\rm rad} - \gamma_{\mu,\rm int})$, from which we see that there will be a real zero at some frequency when $\gamma_{\mu,\rm rad} = \gamma_{\mu,\rm int}$. 
When the radiative loss equals the absorption loss the S-matrix has an eigenvalue equal to zero at a real frequency: one can send in a specific wavefront and it will be indefinitely trapped and hence absorbed.
The multichannel case of this is CPA, and the single-channel case, for which a specific wavefront is not required, is the usual critical coupling to a resonator.
However, as noted, this only corresponds to critical coupling when all of the loss is due to absorption, whereas we will present the full generalization of reflectionless coupling in the next section.

Up to this point we have just reviewed the known properties of the S-matrix in this operator representation. 
Henceforth we are focusing on the zeros of the matrix $\bf R_{\rm in}$, when it differs from the full S-matrix, and seeking the condition for it to have zeros.

We now present results from adapting this formalism to treat R-zeros/RSMs; unlike the results presented in the previous section, these results were not known previous to the derivations in Ref.~\cite{rsm_arxiv}.
The basic approach is to represent the matrix ${\bf R}_{\rm in}$ through appropriate projection operators applied to the S-matrix, and then to perform a similar, but more involved set of manipulations to obtain a relationship between the $\det {\bf R}_{\rm in}$ and a ratio of determinants of wave operators to similar to but distinct from those that determine $\det{\bf S}$.

The result is \cite{rsm_arxiv}:
\begin{equation}
\label{eq:det(R)}
    \det {\bf R}_{\rm in}(\omega)
    = \frac{\det\boldsymbol{(}{\bf A}_0(\omega)-{\boldsymbol \Delta}(\omega)-i[{\boldsymbol \Gamma}_{\rm in}(\omega) - {\boldsymbol \Gamma}_{\rm out}(\omega)]\boldsymbol{)}}{\det{\boldsymbol (}{\bf A}_0(\omega)-{\boldsymbol \Delta}(\omega)+i{\boldsymbol \Gamma(\omega)}{\boldsymbol )}},
\end{equation}
where the only (but crucial) difference from Eq.~\eqref{eq:det(S)} is the replacement of the operator ${\boldsymbol \Gamma}$ by the {\it difference} of two operators associated with the input and output channels respectively: $\boldsymbol \Gamma_{\rm in} - \boldsymbol \Gamma_{\rm out} \equiv \mathbf W_{\rm in}^{\phantom{\dagger}} \mathbf W_{\rm in}^{\dagger} - \mathbf W_{\rm out}^{\phantom{\dagger}} \mathbf W_{\rm out}^{\dagger}$.
Here the subscripts refer to the sectors of the operator ${\bf W}$ introduced previously that connect the discrete states of the scatterer/resonator to the asymptotic incoming channels and outgoing channels (respectively), which were specified in the definition of $\bf R_{\rm in}$.  
Eq.~\eqref{eq:det(R)} is the central mathematical result of our theory of reflectionless states.
Similar to Eq.~\eqref{eq:det(S)}, Eq.~\eqref{eq:det(R)} relates the determinant of the $N_{\rm in}$-by-$N_{\rm in}$ matrix $\bf R_{\rm in} $ to a ratio of wave operator (functional) determinants which describe the discrete but infinite space of eigenvalues of the scatterer/resonator.

    \section{Properties of R-zeros and RSMs}
    \subsection{General Properties \label{sec:general properties}}

From Eq.~\eqref{eq:det(R)} we can draw a number of critical conclusions:
\begin{itemize}

    \item For reasons analogous to those arising from Eq.~\eqref{eq:det(S)}, we can conclude that the matrix $\bf R_{\rm in} $ also has a countably infinite set of zeros and poles at discrete complex frequencies.
    
    \item Because the denominator is the same as in Eq.~\eqref{eq:det(S)}, the poles of $\bf R_{\rm in} $ are identical to those of ${\bf S}$ (excluding certain non-generic cases). 
    
    \item However the {\it zeros} of $\bf R_{\rm in}$ (R-zeros) are generically at distinct frequencies in the complex plane from those of ${\bf S}$.
    As noted above, the R-zero spectrum is a new complex spectrum with a distinct physical meaning from S-matrix zeros or poles \cite{2018_Bonnet-BenDhia_PRSA,rsm_arxiv}.
    
    \item Because the operator in the numerator of Eq.~\eqref{eq:det(R)} is not the hermitian conjugate of that in the denominator, even when the scatterer is passive (no loss or gain) the R-zeros are not the complex conjugate of the poles and can appear in either the upper or lower half-plane without the addition of loss or gain.
    In particular, a lossless scatterer can have a real R-zero (RSM), although generically this requires parameter tuning. 
    
    \item The fact that the positive semidefinite coupling operator $\mathbf W_{\rm in}^{\phantom{\dagger}} \mathbf W_{\rm in}^{\dagger}$ gives a contribution to the numerator of Eq.~\eqref{eq:det(R)} of opposite sign to that of the outgoing channels  (which has the usual sign for the S-matrix) implies that qualitatively the incoming channels function as a kind of ``radiative gain" for the R-zeros, whereas the outgoing channels function qualitatively in the usual manner as radiative loss.
    Heuristically we can expect that an R-zero can become an RSM when the total coupling in from the input channels balances the total coupling out from the radiative channels, or, if there is loss or gain in the resonator itself, when all of these terms are balanced  to cancel.
    If we are in the regime of isolated resonances there will be only one set of relevant couplings to balance and this can be regarded as a multichannel generalization of critical coupling. 
    However to excite this RSM one will still need to send in the correct  coherent wavefront obtained from the eigenvalue equation for $\bf R_{\rm in}$.
    
    \item  Our result shows that the full operator in the numerator of Eq.~\eqref{eq:det(R)} will have an infinite number of zeros in general, even if we are not in the regime of isolated resonances.
    In this case there is no single scalar condition for achieving RSM and no meaningful generalization of the critical coupling concept.
    Nonetheless R-zeros can be calculated, and by varying parameters of the scatterer/resonator, can be tuned to RSM. 
    So the existence of reflectionless states, with tuning, is a robust property of electromagnetic scattering and does not require a high-\Q\ resonator.
    An example of tuning to RSM in a low-\Q\ multiwaveguide junction is given in Fig.~\ref{fig:octopus} below.
    
    \item {\bf Tuning R-zeros to RSM}: Although there can be complications from interference between resonances, qualitatively speaking, adding loss to the scatterer will cause the R-zeros to flow downward in the complex plane and gain will cause them to flow upward.
    Similarly, altering the geometry of the scatterer/resonator so as to enhance the coupling of the output channels will cause the R-zeros to move downwards and decreasing the coupling will cause them to move upwards (and vice-versa for the input channels).  
    \hl{In certain cases geometric tuning may affect both input and output channels at once and these couplings may not be separately controlled, making it difficult to tune to RSM.
    However}, in most cases tuning a single structural parameter will be sufficient to allow perfectly reflectionless excitation of the structure; although the correct relative amplitudes and phases of the input channels will be required to access it.  
    
    \item  When the scatterer/resonator has discrete symmetries, in some cases no tuning will be required to find RSMs (real R-zeros).
    Textbook examples are different types of balanced two-mirror resonators, which we will refer to collectively as ``Fabry-P{\'e}rot" (FP) resonators.
    Such resonators have both parity ($\cal{P}$) and time-reversal ($\cal{T}$) symmetry.
    Other more recent examples are one-dimensional photonic structures with balanced gain and loss such that the product of parity and time ($\cal{PT}$)  is preserved.
    Here, unlike the FP resonators, the real RSMs are unidirectional, and can only be accessed from one side or the other.
    We will see in the next section that in both cases the RSMs can be lost due to spontaneous symmetry breaking at an RSM exceptional point (EP).
    
\end{itemize}

    \subsection{Symmetry Properties of R-zeros and RSMs \label{sec:Symmetry Properties}}

The general formulas analyzed above only prove the existence of R-zeros in the complex plane, and therefore, since the real axis has zero measure in the plane, without parameter tuning or symmetry constraints a generic system will have no
RSMs.
However, well known examples, such as the FP resonator, have an infinite number of RSMs, apparently due to symmetry. 
In Ref.~\cite{rsm_arxiv} a detailed analysis is given of the implications of various discrete symmetries on the R-zero spectrum. 
Here we will only present the main conclusions and illustrate them with simple one-dimensional examples.

We will focus on discrete symmetries and their effect on the R-zero spectrum.
The specific symmetries we will analyze here are time-reversal, parity, and the product of the two, as well as the very important case of systems with both symmetries (\cp$+$\ct), exemplified by the FP resonator.
In the context of one-dimensional electromagnetic scattering, \ct\ symmetry requires that the resonator has a real dielectric function, $\varepsilon(x)$; \cp\ symmetry requires $\varepsilon (x) = \varepsilon (-x)$, but need not be real, and \cp\ct\ symmetry  $\varepsilon (x) = \varepsilon^*(-x)$, but again it needs not be real.

    \subsubsection{Time-reversal Symmetry ($\mathcal{T}$) and Parity Symmetry ($\mathcal{P}$) \label{sec:sym_P}}

The time reversal operator (${\cal T}$) complex conjugates the wave equation in the frequency domain. 
If the system has \ct\ symmetry then it maps a Left-incident R-zero to a Right-incident R-zero of the same cavity but at frequency $\omega^*$. 
Hence:
\begin{itemize}

    \item If a cavity with \ct\ symmetry (real $\varepsilon$) has a Left R-zero at $\omega$, then it will have a Right R-zero at $\omega^*$.
    If it is tuned to RSM without breaking \ct\ symmetry then it will then be bidirectional, i.e. it will have a Left RSM and Right RSM at the same frequency.
    
\end{itemize}

The parity operator \cp\ maps $x \rightarrow -x$ in the wave equation; if the dielectric function has parity symmetry then it maps a Left R-zero at $\omega$ to a Right R-zero of the same cavity at the same frequency.
Hence:
\begin{itemize}

    \item All R-zeros of parity symmetric one-dimensional systems are bidirectional, whether or not they are real.  
    
\end{itemize}

\begin{figure}[tb!]
    \centering
    \includegraphics[width=0.9\columnwidth]{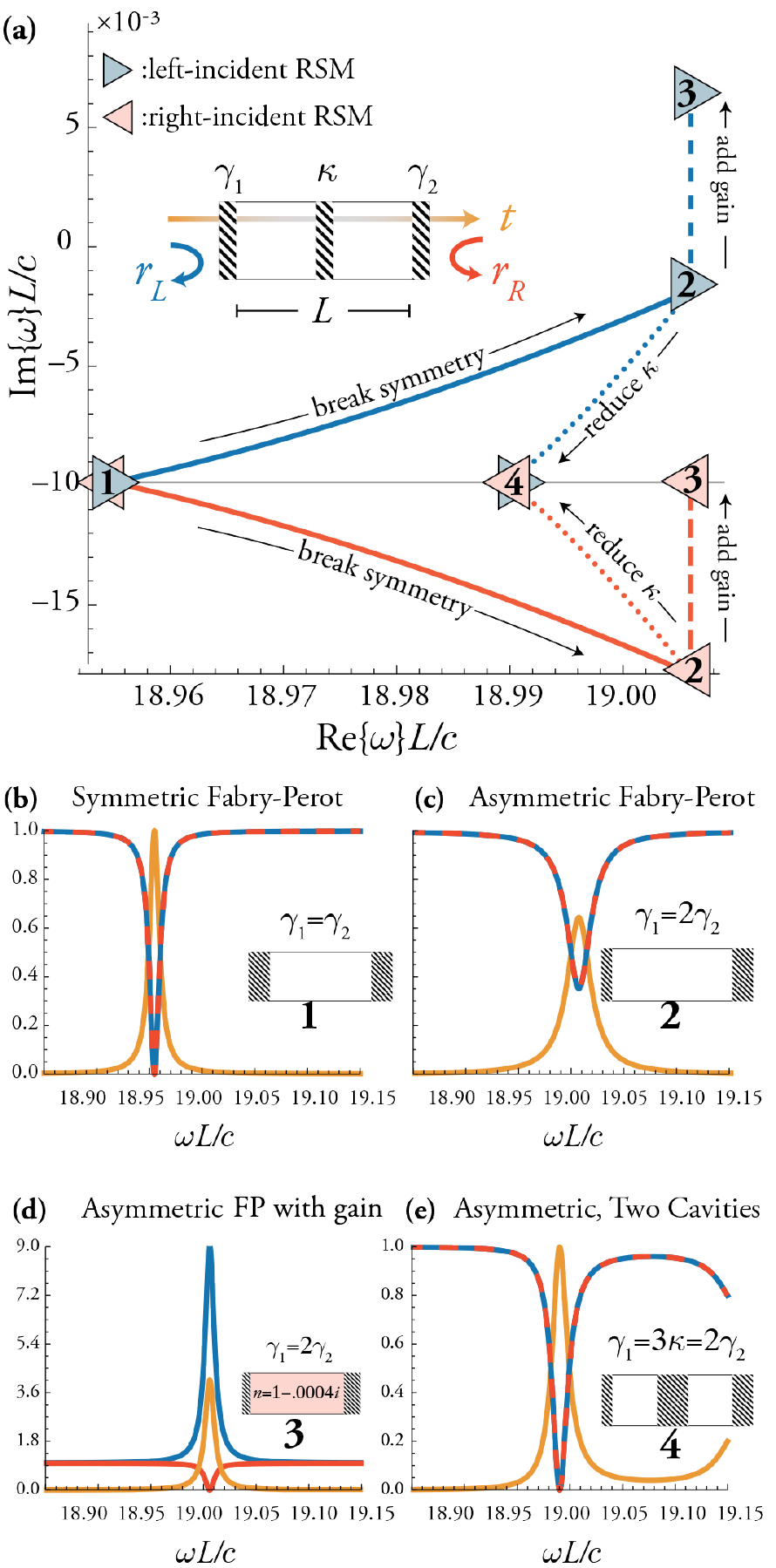}
    \caption{(Color) Illustration of RSMs and R-zero spectrum for simple two- and three-mirror resonators of length $L$ in 1D, consisting of $\delta$-function mirrors of strengths $\gamma_1^{-1}, \gamma_2^{-1}$ and $\kappa^{-1}$, as indicated in the schematic in (a).
    Throughout, we fix $\gamma_2 \equiv c/L$.
    Blue and red lines $1 \to 2$ indicate the effect of breaking symmetry by varying $\gamma_1$ from $\gamma_2 \to 2 \gamma_2$.
    A bidirectional RSM [as in (b)] splits into two complex-conjugate R-zeros off the real axis and a reflectionless steady-state (RSM) no longer exists, as in (c).
    Adding gain to the cavity, indicated by blue and red lines $2 \to 3$, brings the lower R-zero to the real axis (but not the upper one), creating a Right-incident amplifying RSM, as in (d).
    Alternatively, adding a middle mirror and reducing its $\kappa$ from $\infty \to 2 \gamma_2/3$ is sufficient to bring both R-zeros back to the real axis ($2 \to 4$ in (a)), creating simultaneous Left and Right RSMs at a different frequency from the symmetric Fabry-P{\'e}rot resonator (see (e)), without restoring parity symmetry.
    }
    \label{fig:fp}
\end{figure}

Both \cp\ symmetry and \ct\ symmetry map from Left R-zeros to Right R-zeros and imply relationships between these spectra, but neither alone implies R-zeros are real.
Hence in systems with only one of these symmetries, parameter tuning will be required to create RSMs. 
A simple example illustrating this with an asymmetric FP resonator is shown in Fig.~\ref{fig:fp}(a-e).
Here we illustrate tuning to RSM with both \ct-preserving (geometric) tuning and \ct-breaking (gain/loss) tuning.

Starting with a symmetric FP cavity, with \cp$+$\ct\ symmetry, we first break parity symmetry but maintain \ct\ symmetry by simply making the mirror reflectivities unequal (Fig.~\ref{fig:fp}(a) solid lines). 
The Left-incident and Right-incident R-zeros leave the real axis as complex conjugate pairs, as required by \ct\ symmetry, and there are no remaining RSMs (there are resonances, but without zero reflection).
To create a Right-incident RSM we add gain to the system, breaking \ct\ symmetry, with the correct value to bring the Right R-zero in the lower half-plane back to the real axis; at the same time the Left R-zero is pushed further away and the RSM created is unidirectional.
The Left R-zero could also be tuned to RSM by adding an equivalent amount of loss. 
To create a bidirectional RSM, we instead add a third lossless mirror to the resonator and we find that tuning the reflectivity of the new (middle) mirror can bring the system back to real axis. 
Since \ct\ symmetry has been maintained, the RSM must be bidirectional.
Note that in both cases the tuning {\it has not restored parity symmetry}.
All such RSMs are in this sense  ``accidental", achieved by parameter tuning and without an underlying symmetry. 
Geometric tuning is of particular interest for reflectionless states of complex structures where adding loss or gain may not be practical or desirable (see the examples in Figs.~\ref{fig:PT}--\ref{fig:converter} below).

    \subsubsection{\cp\ and \ct\ Symmetry and Symmetry-breaking Transition\label{sec:sym_P_and_T}}

The most prominent example of a system with an infinite number of RSMs was already mentioned above, two-mirror resonators (where ``mirror" includes the many types of reflecting structures used in photonics) which will be referred to collectively as FP resonators.
In fact one can easily check that all R-zeros of an FP resonator are real and in one-one correspondence with the resonances of the structure.  
This example might suggest that any structure with both \cp\ and \ct\ symmetry should have only real R-zeros (RSMs). 
Previous work on multi-mirror resonators \cite{stadt_1985_osa,stone_1990} however has found cases which violate this expectation, but to our knowledge no general reason for this fact or qualitative understanding of it has been given. 
The symmetry analysis we present here provides such a framework, with additional implications which are new.
A more detailed study of this case is in preparation \cite{Sweeney_P_and_T}; here we present a brief outline of the problem, illustrated with a simple example.

\begin{figure}[tb!]
   \centering
   \includegraphics[width=0.85\columnwidth]{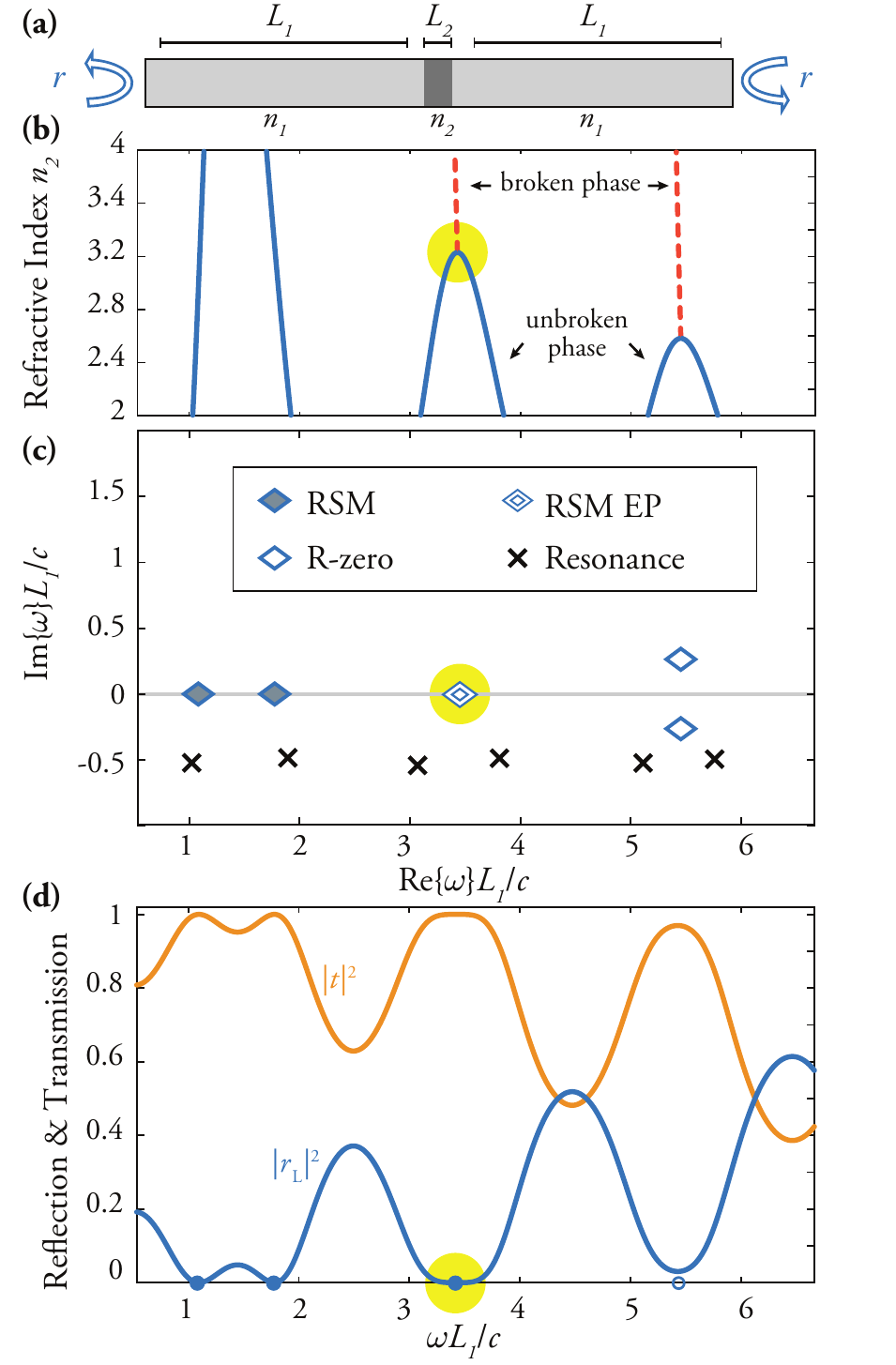}
   \caption{(Color) RSMs in a \cp$+$\ct\ symmetric structure.
   {\bf (a)} Symmetric three-slab heterostructure in air, with refractive index $n_1$ in the outer sections, which are of length $L_1$, and variable index $n_2$ in the middle, of length $L_2$; here $n_1=1.5$ and $L_2=.15L_1/n_2$.
   {\bf (b)} Real part of the R-zero frequencies as the central index $n_2$ is increased.
   For small $n_2$, the R-zeros are real-valued RSMs and in the unbroken phase (solid blue lines), while for large $n_2$, some R-zeros have entered the broken phase (red dashed).
   After two RSMs meet at an RSM EP, they split into two R-zeros at complex-conjugate frequencies as $n_2$ is further increased.
   {\bf (c)} Spectra of the R-zeros/RSMs and resonances in the complex-frequency plane at $n_2 = 3.23$ where two (bidirectional) RSMs meet at two degenerate EPs (one for Left RSMs and one for Right RSMs).
   {\bf (d)} Reflection and transmission spectra for the same $n_2$ as in (c); blue filled dots mark the RSM frequencies, open blue dot is real part of complex R-zero, which has already entered the broken phase.
   Yellow highlight in (b--d) indicate the same RSM EP, which exhibits quartically-flat reflection and transmission.
   }
   \label{fig:PT}
\end{figure}

First we must draw your attention to an important point.
Even if the structure/resonator has both \cp\ and \ct\ symmetry, the wave operator for the R-zero spectrum does not. 
As already noted, both the \cp\ and \ct\ operations map Left R-zeros back to Right R-zeros; hence the boundary conditions are not invariant under these operations.
However it is easy to confirm that if a structure has both \cp\ and \ct\ symmetry (including the asymptotic regions), then the product \cp\ct\ maps the R-zero spectrum for a single directionality back to itself as follows:
\begin{itemize}  

    \item If a cavity has both \cp\ and \ct\ symmetry, its R-zeros occur in complex- conjugate pairs or are non-degenerate and real (i.e.~RSMs).
    The frequency spectrum of the Left R-zeros and the Right R-zeros are the same, so that all RSMs and R-zeros are bidirectional.

\end{itemize}

Since both the case of real RSMs without tuning, and complex-conjugate pairs of R-zeros are allowed by symmetry, we expect both cases to occur in some structures. 
We are already familiar with FP resonators for which all of the R-zeros are RSMs; in Fig.~\ref{fig:PT} we examine the simplest example beyond the FP case for which not all R-zeros are RSMs.
Similarly to the example of Fig.~\ref{fig:fp}, we add a middle mirror to the FP resonator, but now the initial FP resonator has balanced mirrors and \cp$+$\ct\ symmetry. 
It is well-known that such a coupling mirror will create two coupled cavities for which the original resonances will be paired up as quasi-degenerate symmetric and anti-symmetric doublets with twice the original free spectral range.
However the doublets cannot become fully degenerate until the internal mirror becomes totally opaque and we simply have two separate identical one-sided cavities.
In contrast there are continuity arguments which we omit here that imply that the RSMs must disappear at a finite coupling. 
In other words, as the coupling mirror becomes more opaque there is a finite coupling at which a pair of RSMs associated with a resonant doublet meet, and then leave the real axis as complex conjugate R-zeros. 
This is an example of a spontaneous \cp\ct\ symmetry-breaking transition~\cite{2017_Feng_nphoton,2018_ElGanainy_nphys,2019_Miri_Science,2019_Ozdemir_nmat} driven by the increase of the coupling mirror opacity.
{\it Thus we have shown the existence of a \cp\ct\ transition in a lossless system}.
This is the first such example to our knowledge, and it is possible because the R-zero boundary conditions themselves are non-hermitian, even if the differential operator in the wave equation doesn't have a complex dielectric function.
This behavior of the three-mirror lossless resonator is illustrated in Fig.~\ref{fig:PT}.

Finally, when the two RSMs meet on the real axis we have a degeneracy of the non-hermitian R-zero eigenvalue problem, so this must correspond to an exceptional point of the underlying wave operator, which in this case happens on the real axis, and not in the complex plane as happens for many other studied cases for which two resonances become degenerate.
More precisely, there are two degenerate EPs when bidirectional RSMs meet; one for the Left RSM spectra and one for the Right RSM spectra.
For a second order EP such as this, on the real axis, previous work has shown~\cite{2019_Sweeney_PRL} that the associated lineshape is altered to a quartic flat-top shape; this behavior is visible in Fig.~\ref{fig:PT}.
The existence of this specific tuned behavior of the three mirror resonators was predicted long ago~\cite{stadt_1985_osa,stone_1990} and is used in designing ``ripple-free" filters. 
Such filters are the analogs of Butterworth filters in electronics~\cite{butterworth_1930}.
However the previous work does not seem to have identified this as an exceptional point, associated with a \cp\ct\ transition. 
A much more in depth analysis of the physics of this transition will be given elsewhere  \cite{Sweeney_P_and_T}.

    \subsubsection{Parity-Time symmetry (${\cal PT}$) and Symmetry-Breaking Transition\label{sec:sym_PT}}

Reflectionless states have been previously studied extensively~\cite{2011_Lin_PRL,2012_Ge_PRA} for one-dimensional resonators which have only \cp\ct\ symmetry, but not \cp\ and \ct\ separately; in this case the condition $\varepsilon(x) = \varepsilon^*(-x)$ holds, but $\varepsilon$ is not real (balanced gain and loss).
This case is discussed in detail in \cite{rsm_arxiv}; it is straightforward to show that in this case R-zeros are either real or come in complex conjugate pairs, but there is no connection between the Left and Right R-zero spectra.
Hence all R-zeros and RSMs in this case are unidirectional.
If one starts with a system with \cp\ and \ct\ symmetry such as the standard FP resonator, and now adds balanced gain and loss to maintain \cp\ct\ symmetry, again one will see pairs of RSMs move towards each other on the real axis, pass through an EP, before emerging as complex conjugate pairs of R-zeros.
This example, given in Ref.~\cite{rsm_arxiv}, then differs from the example shown in Fig.~\ref{fig:PT} only in that the Left R-zeros and Right R-zeros no longer move together.

\section{Applications of R-zero/RSM theory}
    \subsection{Relationship to Coupled Mode Theory\label{sec:RSM_TCMT}}

The preceding results were derived directly from Maxwell's equations and involve no approximation.
In many circumstances an approximate analytic model will be adequate and desirable for simplicity.
In photonics a standard tool is the temporal coupled-mode theory (TCMT)~\cite{Haus_book, 2003_Fan_JOSAA, 2004_Suh_JQE, 2018_Wang_OL, 2019_Zhao_PRA, 2017_Alpeggiani_PRX}, which is a phenomenological model widely used in the design and analysis of optical devices~\cite{2012_Verslegers_PRL, 2014_Peng_nphys, 2014_Hsu_NL, 2015_Zhen_Nature}.
The TCMT formalism is derived from symmetry constraints~\cite{Haus_book, 2003_Fan_JOSAA, 2004_Suh_JQE, 2018_Wang_OL, 2019_Zhao_PRA} rather than from first principles, yet it leads to an analytic relation between the determinant of the scattering matrix and the underlying Hamiltonian that is similar to Eq.~\eqref{eq:det(S)} and is reasonably accurate in many cases.
The appropriate comparison between TCMT theory and the exact RSM theory presented here is given in Ref.~\cite{rsm_arxiv}.
Here we will only quote one relevant result of that analysis.
Not surprisingly, it is possible to adapt the TCMT analysis which leads to an expression for the S-matrix in terms of an ``effective Hamiltonian", so as to find an expression for the ${\bf R}_{\rm in}$ matrix, and a condition for it to have an R-zero.
In this expression, as in our Eq.~\eqref{eq:det(R)}, the coupling coefficients for the input channels to the resonator appear as an effective gain term and coefficients for the output channels appear as an effective loss term.
Very often in the TCMT formalism the single-resonance approximation is used, in which the effective Hamiltonian is replaced by a number equal to the complex energy of the resonance.
When a similar approximation is used for ${\bf R}_{\rm in} $ to determine the R-zeros one finds~\cite{rsm_arxiv}:
\begin{equation}
\begin{aligned}
\label{eq:omega_RSM_single_mode}
    &\omega_{\rm RZ} = \left( \omega_0-i\gamma_{\rm nr} \right) + i \left( \gamma_{\rm in} - \gamma_{\rm out} \right), \\
    &\gamma_{\rm in} \equiv \sum_{n \in F}|d_n|^2/2,
    \quad \gamma_{\rm out} \equiv \sum_{n \notin F}|d_n|^2/2,
\end{aligned}
\end{equation}
here $\omega_{\rm RZ}$ is the R-zero frequency, $\omega_0$ is the real part of the resonance frequency, $d_n$ is the coupling coefficient (partial width) of the mode to the $n$-th radiative channel, $\gamma_{\rm in},\gamma_{\rm out}$ are then the total radiative rates in and out respectively for the resonance, and $\gamma_{\rm nr}$ is the non-radiative rate associated with loss or gain in the resonator.
An RSM arises when the various imaginary terms for $\omega_{\rm RZ}$ cancel, and the structure is illuminated with the corresponding wavefront, determined by the eigenvector of ${\bf R}_{\rm in} $.  
This then corresponds to the critical coupling condition, generalized to multichannel inputs and outputs.
Here we have used the standard notation in TCMT theory for the partial coupling rates into or out of a given channel.
In our theory, evaluating the matrices
${\bf W}^{\phantom{\dagger}}_{\rm in} {\bf W}_{\rm in}^{\dagger}, {\bf W}^{\phantom{\dagger}}_{\rm out} {\bf W}_{\rm out}^{\dagger}$, in the limit of a single resonance yields a similar critical coupling relationship.

Hence TCMT theory within the single (high-\Q) resonance approximation gives an analytic basis for the concept of generalized critical coupling introduced above, and also shows its limitations. 
One implication of this result is that R-zeros with different numbers of input and output channels will have the same ${\rm Re}\{\omega\}$ as the underlying resonance and will be simply shifted vertically in the complex plane along a line between the S-matrix pole (resonance) and its zero.
The appropriate input wavefront for the RSM will just be that of the outgoing resonance, phase conjugated for the chosen input channels.
This behavior is found in the example shown in Fig.~\ref{fig:octopus}a, for the case of a high-\Q\ resonator.
Conversely, the TCMT single resonance result fails for a more open resonator (Fig.~\ref{fig:octopus}b), where multiple resonances mediate the scattering within the resonator.
Nonetheless, the exact R-zero/RSM approach can be used to tune to RSM numerically.

The single-resonance scenario is the simplest example of an R-zero, yet it already is sufficient to explain the impedance-matching conditions previously found using TCMT in waveguide branches~\cite{2001_Fan_JOSAB}, antireflection surfaces~\cite{2014_Wang_Optica}, and polarization-converting surfaces~\cite{2017_Guo_PRL}.

    \subsection{Reflectionless States in Complex Structures: Examples}

Here we show two examples of RSMs engineered in complex photonic structures; these are the kinds of impedance-matching problems it would be very difficult to solve without our theory and associated computational approaches. 
As noted above, R-zero spectra can be calculated by a modified PML method in many cases, and by a boundary matching method in all cases.  
The first of the examples below was done by the PML method, which is somewhat simpler.
The second was done by the matching method and could not be done with PMLs, because the input and output channels are not separated in the asymptotic regions.
We will discuss the solution methods very briefly in the next section.

\begin{figure}[tb!]
    \centering
    \includegraphics[width=0.85\columnwidth]{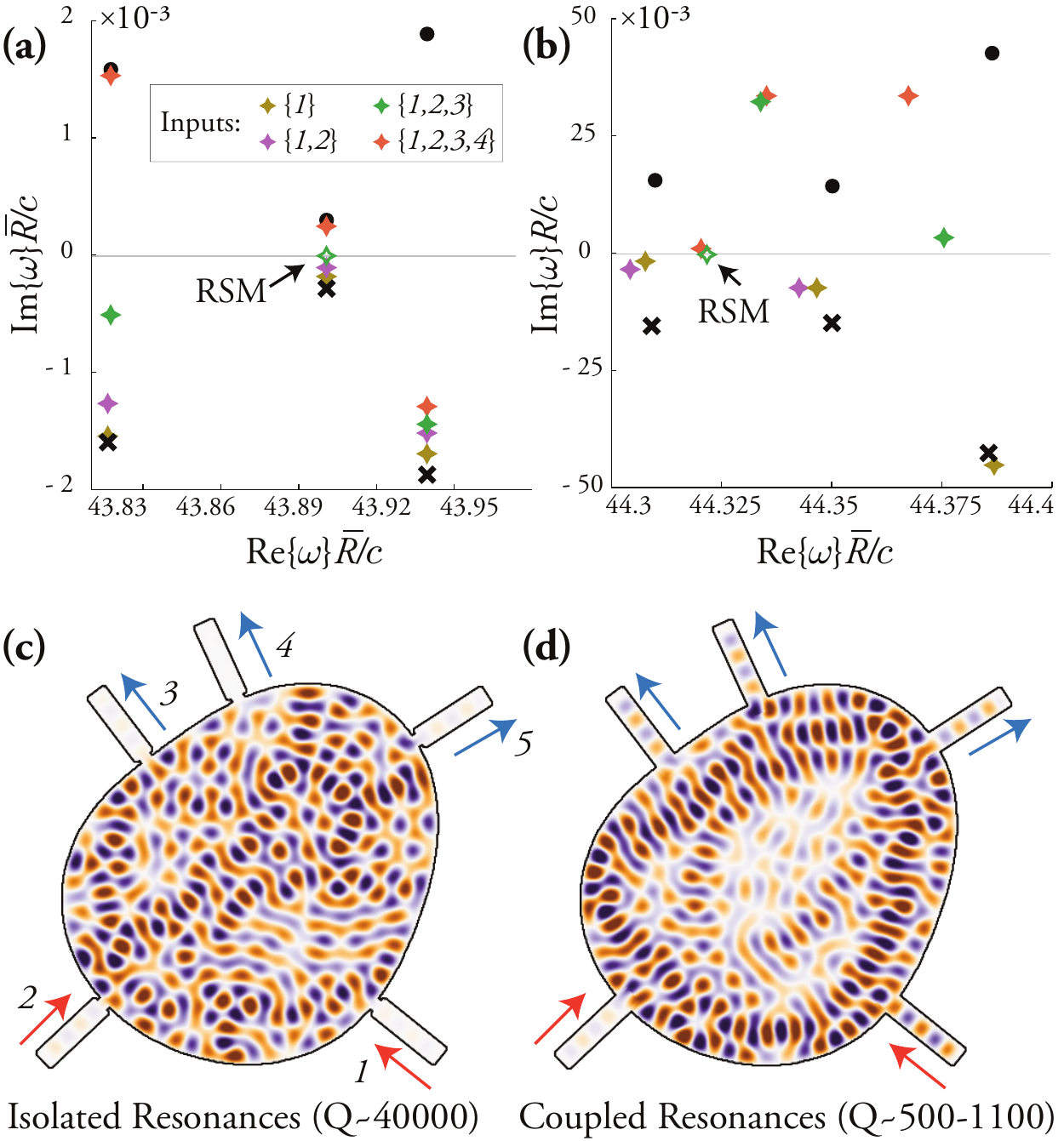}
    \caption{(Color) Asymmetric lossless waveguide junction/resonator (mean radius $\bar R$) coupled to five single-mode waveguides, with constrictions at the ports to the junction.
    {\bf (a)} Numerically calculated R-zero spectrum for a weakly coupled, high-\Q\ junction with well isolated resonances.
    Black x and dot are purely outgoing (resonance) and incoming (S-matrix zero) frequencies, which are complex conjugates.
    Colored stars are R-zeros for various choices of input channels; the legend indicates which channels are inputs, with the channel labels given in (c). The R-zeros cluster vertically above the resonance frequency and below the S-matrix zero frequency, as predicted by single-resonance TCMT approximation Eq.~\eqref{eq:omega_RSM_single_mode}.
    The common width of the constrictions for waveguides $\{4,5,6\}$ is slightly tuned to make a 2-in/3-out R-zero real, creating an RSM.
    {\bf (b)} R-zero spectrum for the same junction but with the constrictions opened, which lowers the \Q\ of the resonances (note change in vertical scale). The linewidths of the resonances are now comparable to their spacing. 
    Due to multi-resonance effects, the R-zeros are spread out along the real and complex frequency axis and are no longer associated with a single resonance.
    Nonetheless, by slightly tuning the constriction width as before, a 2-in/3-out R-zero is again made real (RSM), as in the high-\Q\ case.
    {\bf (c,d)} The mode profiles of the RSMs for the high-\Q\ (c) and low-\Q\ (d) cases.
    }
    \label{fig:octopus}
\end{figure}

In Fig.~\ref{fig:octopus}(a) we show results for an asymmetric cavity much larger than the wavelength of the exciting radiation, with a smooth boundary connected to five single-mode waveguides, without any internal gain or loss.
Here and in the next example we are only going to use geometric/index parameter tuning to achieve a flux-conserving RSM.
The structure has no discrete symmetries, so there is no reason that there should exist any RSMs for such structures without parameter tuning.
In addition such cavities are well known to have many pseudo-random wave-chaotic states, so that the resonances do not have any simple spatial structure or ray orbit interpretation, making intuitive design approaches to generating the appropriate interference behavior impossible. 
We consider this structure in two limits: (a) the limit in which the waveguide ports are pinched off by constrictions to create a high-\Q\ cavity and large non-resonant reflection at the ports, and (b) the limit of essentially open ports with slight width tuning, for which the cavity has much lower \Q\ and multiple resonances participate in scattering.

In the case (a) the behavior is as predicted by the single-resonance approximation, discussed just above.
R-zeros are lined up vertically in the complex plane between the pole and the zero, and one of them (a two-in, three-out case) has been tuned to the real axis. 
The internal field (real part shown in (c)) is chaotic looking and coincides well with the single resonance associated with these R-zeros.

In the case (b), where we impose the same R-zero boundary conditions, we see very different behavior of the R-zero spectrum, characteristic of transmission through multiple resonances.
The R-zeros are spread out in the complex plane and do not lie on a line coincident with any one resonance, nor is the input wavefront or internal field associated with a single resonance. 
Tuning to RSM is achieved by a very slight variation in the width of one of the outgoing waveguides at the port.
Further details are given in the figure caption.

\begin{figure}[tb!]
    \centering
    \includegraphics[width=0.85\columnwidth]{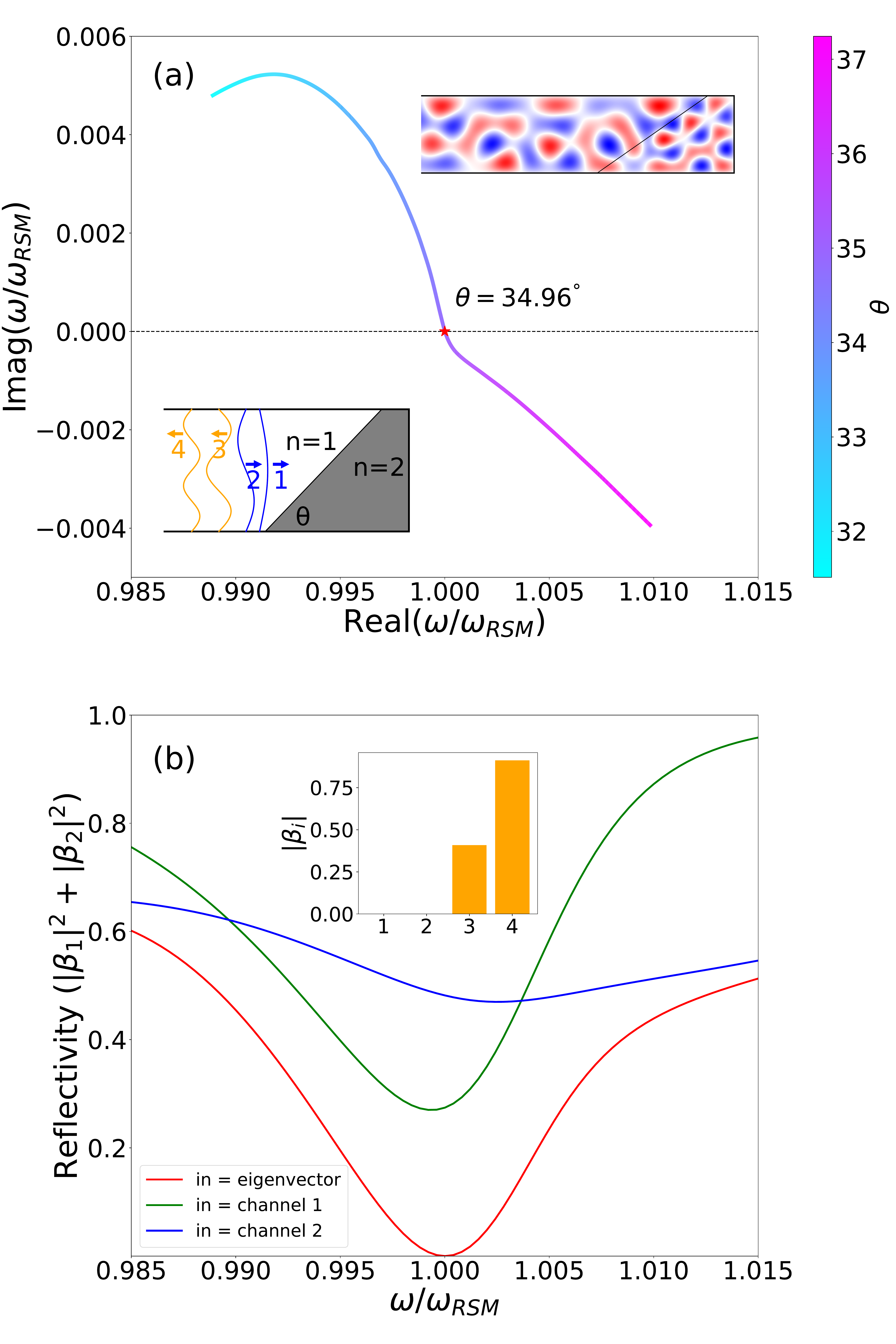}
    \caption{(Color) Illustration of an RSM in a four-mode waveguide acting as a mode-converter from a superposition of input waveguide modes 1 and 2 into output modes 3 and 4.
    {\bf (a)} Trajectory of R-zero in the complex-frequency plane as wedge-angle $\theta$ is tuned.
    The R-zero crosses the real axis at $\theta = 34.96^\circ$ (red star), becoming an RSM. 
    Insets show a schematic of the structure and the real part of the RSM field profile.
    {\bf (b)} Reflectivity into modes 1 and 2 with different incident wavefronts, for $\theta = 34.96^\circ$.
    Red curve has input $\alpha_0(\omega)$, defined to be the eigenvector of ${\bf R}^\dagger_{\rm in}(\omega){\bf R}_{\rm in}(\omega)$ with the smallest eigenvalue.
    ${\bf R}_{\rm in}(\omega)$ is the $2\times2$ upper left block of the scattering matrix.
    The incident wave $\alpha_0(\omega_{\rm RSM})=[0.7982, -0.5642+0.2158i, 0, 0]$  generates the reflectionless output $\beta(\omega_{\rm RSM})=[0, 0, 0.1767-0.3689i, 0.7682+0.4925i]$.
    Meanwhile, the inputs from only mode 1 or only mode 2 (green and blue curves) have non-zero reflectivity for all frequencies.
    The inset shows the output amplitude $|\beta(\omega_{\rm RSM})|$ for the eigenvector input $\alpha_0(\omega_{\rm RSM})$.
    }
    \label{fig:converter}
\end{figure}

The second example, shown in Fig.~\ref{fig:converter}, is a structure which functions as a lossless mode converter in reflection.
It is a multimode empty waveguide terminated by an angled wedge of purely real index $n=2$ material and then by a perfectly reflecting wall. 
In the case where the waveguide has only two modes it is relatively simple to find a wedge angle which converts, e.g., mode one into mode two perfectly after reflection, and a simple Fresnel scattering analysis could be used to find the necessary angle  to a good approximation. 
Here however the waveguide has four propagating modes and the R-zero problem is to use modes one and two as inputs and modes three and four as outputs.
Our general theory implies that such a solution should exist at some $\omega$ as one tunes a single parameter such as the wedge angle.
Indeed, Fig.~\ref{fig:converter}(a) shows that as the wedge angle is tuned, one of the R-zeros crosses the real axis and becomes an RSM.  

As there are multiple input channels, zero reflection is achieved only when the correct superposition of modes one and two is used.
The smallest eigenvalue of ${\bf R}_{\rm in}^\dagger(\omega) {\bf R}_{\rm in}^{\phantom{\dagger}}(\omega)$ gives the smallest possible reflectivity, and Fig.~\ref{fig:converter}(b) shows that this vanishes at $\omega_{\rm RSM}$, while the reflected intensities for single-channel inputs do not anywhere in the vicinity of $\omega_{\rm RSM}$.

Finally, in Ref.~\cite{rsm_arxiv} an example was presented of an RSM solution to a different and challenging problem in free-space scattering: designing a dielectric antenna much larger than the input wavelength such that it perfectly reflects a monopole input signal (assuming 2D scalar waves) into higher multipoles in the scattered field. 
Although here we considered scalar waves, the method can be straightforwardly generalized to vector electromagnetic waves.
This additional example, not involving waveguides or mirror resonators, illustrates the versatility of the R-zero/RSM theory for electromagnetic design.

Finding these types of impedance-matched solutions for open multi-channel structures would be difficult without our theory, nor were there, to our knowledge, previous formulations which guarantee a solution exists for the appropriate input wavefront with single parameter tuning.
As already noted, the usual critical coupling concept does not extend to these type of low-\Q\ structures.

    \subsection{Solution Method for R-zero/RSM problems\label{sec:RSM_wave_operator}}

We make a few brief remarks on solving the R-zero/RSM problem numerically, which will be necessary for essentially all cases of interest.
In principle one could find R-zeros by constructing the full S-matrix of the problem, and then the relevant ${\bf R}_{\rm in} (\omega)$ and search for the zeros of its determinant in the complex plane.
Doing so, however, has numerical disadvantage because ${\bf R}_{\rm in} (\omega)$ changes rapidly with frequency near resonances, which generally makes it harder for such root finding and for other nonlinear eigenproblem solvers. 
Our theory demonstrates that the R-zeros can be found directly from imposing boundary conditions of the wave operator on the boundary of a computational cell, without constructing the scattering matrix in the complex plane.
This typically more efficient procedure then yields the R-zero spectrum in a given frequency range for an initial structure, which typically contains no RSMs. 
However the fact that the R-zeros are eigenvalues of a well-behaved wave operator means that its eigenvalues will move continuously with small changes in the real or imaginary part of the dielectric function. 
Moreover the general form of Eq.~\eqref{eq:det(R)} has given us the intuition to know what kinds of geometric or loss/gain perturbations will move the R-zeros up or down in the complex plane (e.g increasing the coupling of an outgoing channel will tend to move all R-zeros down and vice-versa for an incoming channel). 
While the simple critical coupling picture is not always valid, the tuning of an R-zero to the real axis is by no means a random search in a parameter space. 
In addition, obviously, many different tunings will work if any real frequency in a given range is acceptable.
The upshot is that numerically, engineering a single RSM can be achieved by a small number of iterations of the initial calculation.
Hence, finding the RSMs in many cases is computationally no more difficult than iterating a resonance calculation of the type available in packages such as COMSOL over a number of weakly perturbed structures.

However there is one important difference between resonance calculations and RSM calculations; all resonance calculation can be done using the perfectly matched layer (PML) method since all the asymptotic channels satisfy the same outgoing boundary conditions.
When the system has negligible dispersion, the PML method turns a non-linear eigenvalue problem into an linear one, for which the boundary condition is independent of $\omega$, and this makes the solution easier~\cite{lalanne_2018}.
There also exist conjugated PMLs~\cite{2018_Bonnet-BenDhia_PRSA}, though they are less well-known, which can also implement purely incoming boundary conditions. 
However there is (as yet) no PML method to solve problems, such as the mode converter example above, for which the incoming and outgoing channels overlap in space.
Here we need  to impose matching conditions outside the surface of last scattering based on exactly the set of incoming and outgoing channels chosen for the R-zero problem (this is what we mean by R-zero boundary conditions). 
Those conditions do depend on the frequency, leading to the more complicated non-linear eigenvalue problem one avoids with PMLs.  
However such matching has been done successfully for complex structures in the wave-chaotic regime \cite{tureci_prog_optics_2005} and in ab initio laser theory 
\cite{2008_Tureci_Science},  and we have used similar methods here and in \cite{rsm_arxiv} to solve the mode converter and multipole converting antenna examples.
Even for these more challenging cases the computations remained quite tractable.
More details about the two methods and a derivation of the matching method are given in \cite{rsm_arxiv}.

    \section{Summary and Outlook}

This paper outlines a general theory of reflectionless excitation or impedance-matching of linear waves to finite structures of arbitrary geometry in any dimension, focusing on the case of classical electromagnetic waves and building on the full theoretical framework presented in Sweeney, et al.~\cite{rsm_arxiv}.
The basic framework applies as well to acoustic and other linear classical waves, and to some quantum scattering problems as well.  Because every impedance-matching problem can be posed as the solution of an electromagnetic eigenvalue problem with certain overdetermined boundary conditions at infinity, similar to the problem of finding resonances, it is possible to completely specify necessary and sufficient conditions for solutions to exist.
It is shown that an infinite number of unphysical solutions always exist at discrete complex frequencies, and any single solution can be tuned to the real axis, typically by varying a single parameter of the structure, so as to become a physically realizable steady-state harmonic solution.
These solutions are accessible if it is possible to generate the appropriate input wavefront, which can be determined from the solutions.
We refer to wave solutions of this type as Reflectionless Scattering Modes (RSMs) because they are steady-state solutions adapted to the particular structure and which specify the behavior of each scattering channel at infinity, similar to lasing modes.

While we have an analytic framework for determining the RSMs, the resulting equations will usually need to be solved numerically and we present two methods for doing so which are computationally tractable by adapting standard tools of computational photonics.
Since reflectionless excitation of fairly complex structures is often a goal in photonics, we  believe our theory and approach shows promise for micro and nanophotonic design.
The theory may clarify which design goals are guaranteed an exact solution and which ones are not.
For example, if one has a three-waveguide junction of some geometry, one is guaranteed to be able to find a design for which excitation from waveguide one into waveguides two and three is reflectionless, typically by tuning a single geometric parameter.
However there is no guarantee that further tuning parameters will find a solution which scatters only into waveguide two in some frequency range.
However, since there will be many ways to tune to the reflectionless state of waveguide one, it is interesting to propose a search in this parameter space for the way which minimizes the output into waveguide three, which could be a much more efficient search than an ab initio combinatoric or machine learning-based search of a huge space of structures, most of which are not close to reflectionless.
There are indications in our current results that such an RSM calculation followed by an optimization can succeed.
In this manner we hope that our theory of RSMs can be combined with modern optimization methods to achieve efficiently important design goals in photonic structures.

\begin{acknowledgments}
W.R.S. and A.D.S. acknowledge the support of the NSF CMMT program under grant DMR-1743235.  
We thank Yidong Chong for allowing us to adapt Fig.~2 from his random CPA animation and Abhinava Chatterjee for initial explorations of the \cp$+$\ct\ symmetry-breaking transition.  
\end{acknowledgments}

\bibliography{references}

\end{document}